\documentclass[11pt,pra,aps,onecolumn,superscriptaddress,floatfix,notitlepage,nofootinbib]{revtex4-2}
\usepackage[latin1]{inputenc}
\usepackage{hyperref}
\hypersetup{
    colorlinks=true,
    linkcolor=blue,
    filecolor=magenta,      
    urlcolor=cyan,
    pdftitle={Complementary Noise in Radiation Fields},
    pdfpagemode=FullScreen,
    }
\usepackage{mathrsfs,dsfont}
\usepackage{amsmath}
\usepackage{amsfonts,natbib}
\usepackage{amssymb}
\usepackage{bm}
\usepackage[pdftex]{graphicx}

\usepackage{array}

\usepackage{textpos}
\usepackage{bbold}
\usepackage{adjustbox}

\usepackage{textpos}
\usepackage{bbold}
\usepackage{adjustbox}

\usepackage[normalem]{ulem}

\usepackage{amsmath}
\usepackage{graphicx}

\begin{document}
	\title{Complementary Probes of Gravitational Radiation States} 
\author{Sreenath K. Manikandan}
\affiliation{Nordita, Stockholm University and KTH Royal Institute of Technology, Hannes Alfv\'{e}ns v\"{a}g 12, SE-106 91 Stockholm, Sweden}
\author{Frank Wilczek}
\affiliation{Department of Physics, Arizona State University, Tempe, Arizona 25287, USA}
\affiliation{T. D. Lee Institute, Shanghai 201210, China}
\affiliation{Wilczek Quantum Center and Department of Physics and Astronomy, Shanghai Jiao Tong University, Shanghai 200240, China }
\affiliation{Department of Physics, Stockholm University, AlbaNova University Center, 106 91 Stockholm, Sweden}
\affiliation{Nordita, Stockholm University and KTH Royal Institute of Technology, Hannes Alfv\'{e}ns v\"{a}g 12, SE-106 91 Stockholm, Sweden}
\affiliation{Center for Theoretical Physics, Massachusetts Institute of Technology, Cambridge, Massachusetts 02139, USA}
\date{\today}
   \begin{abstract}
   We demonstrate that the statistical fluctuations in resonant radiation detectors operating in homodyne and heterodyne modes offers additional, complementary information to that obtained from their direct operation as click detectors.  We use this to refine tests of the coherent state hypothesis of interest in connection with gravitational wave fields. 
   \end{abstract}
\maketitle

\section{Introduction}

Experiments over the last century have revealed that nature obeys the laws of quantum mechanics to great accuracy in many circumstances. But few of these experiments significantly involve gravity, and thus far no observed phenomenon in gravity has required elucidation using its (presumed) quantum dynamics.   The relatively recent observation of gravitational radiation ~\cite{abbott,abbott2017gw170817} plausibly opens new possibilities in this regard, since it gives direct access to the gravitational quanta - in great abundance!  It has been suggested that statistical fluctuations in gravitational wave signals might be promising in this regard ~\cite{ParikhPRL,ParikhPRD,ParikhEPJD,tobar_detecting_2024,Manikandan_Wilczek_acoherence}.

Classical or semiclassical calculations of the interactions between matter and gravitational radiation implicitly assume a coherent state description for the gravitational radiation field.  For some -- but not all -- analogous problems in electrodynamics, this can be justified formally using so-called the optical equivalence theorem~\cite{Sudharshan}.  Thus, phenomena that indicate deviations from the {\it coherent state hypothesis\/} for the gravitational radiation field would force us away from that classicality assumption~\cite{Manikandan_Wilczek_acoherence} and require that we bring in a richer quantum description.   
 
Recently ~\cite{Manikandan_Wilczek_acoherence} we suggested that the statistics of resonant bar detectors, when they function as quantized ``click'' detectors 
~\cite{tobar_detecting_2024} are promising in this regard. Here, we extend that work to consider interferometric quantum phase measurement strategies using resonant detectors.  These methods access additional information that is complementary, in the sense of quantum theory, to ``click'' detection. 

 Discrimination of sub-Poissonian quantum states of gravitational radiation is especially challenging ~\cite{CarneyPRD,Manikandan_Wilczek_acoherence,carney_comments_2024}. We find that use of homodyne techniques, which bring in phase sensitivity, and heterodyne techniques, which enhance selected quadratures, can be helpful in this regard. 
 
Our proposed tests of acoherence rely on the foundational work of Ref.~\cite{tobar_detecting_2024}.
There it was suggested that a stimulated absorption probability of order unity can be achieved for gravitational radiation using challenging, but feasible, parameters for resonant bar detector designs. 
The stimulated absorption rate is,
$\gamma_s = \gamma_0 \langle a^\dagger a\rangle $ in the single-mode approximation for the gravitational radiation field with annihilation operator $a$. Here $\gamma_0$ is the spontaneous emission rate. The spontaneous emission rate ($\gamma_0$) and stimulated absorption rate for single gravitons ($\gamma_s$ for the $0\rightarrow 1$ transition) can be estimated for the fundamental acoustic mode of resonant mass detectors to be~\cite{tobar_detecting_2024},
\begin{equation}
   \gamma_0 = \frac{8GML^2\omega^4}{\pi^4 c^5},~~\langle a^\dagger a\rangle  = \frac{h_0^2c^5}{32\pi \hbar G\omega^2},~~\text{and}~~\gamma_s = \gamma_0 \langle a^\dagger a\rangle =\frac{h_0^2ML^2\omega^2}{4\pi^5\hbar},
\end{equation}
where $G$ is Newton's constant, $M,~L$ are the mass and length of the resonant mass detector, and $c$ is the speed of sound, and $h_0$ is the amplitude of the incoming gravitational radiation field. Here $\omega$ is the resonant frequency of the bar detector, $\omega = \pi v_s/L$ for the fundamental acoustic mode of a Weber bar, where $v_s$ is the speed of sound for the material used. Hence the stimulated absorption rate can be re-expressed in terms of material parameters as $ \gamma_s = \frac{h_0^2Mv_s^2}{4\pi^3\hbar}.$

When one estimates for typical Weber bars as was done in Ref.~\cite{tobar_detecting_2024}, $\gamma_0\sim 10^{-33}\text{Hz}$. This can be compared to the spontaneous emission rate for single gravitons from a single Hydrogen atom, that is approximately $10^{-40}\text{Hz}$~\cite{weinberg_gravitation_1972,rothman_can_2006,dyson_is_2013}. 
In comparison, the resonant frequencies for such bars can also be in the kilohertz range, where we know that gravitational waves exist with observable strain amplitudes $h_0\sim 10^{-22}~\text{to}~10^{-21}$.  
For $\omega/(2\pi)~\text{in range } 100\text{Hz to}~200$Hz, this represents $\langle a^\dagger a\rangle \sim 10^{36}$ gravitons ~\cite{dyson_is_2013}.  Therefore, a stimulated absorption probability $\gamma_s\Delta t\sim \gamma_0\langle a^\dagger a\rangle \Delta t$ of order unity can be achieved in practice within the millisecond timescales ($\Delta t\sim 1\text{ms}$) that are relevant to the experiment \cite{tobar_detecting_2024}. 

The product $\gamma_0\Delta t$ represents the graviton-to-phonon conversion efficiency. 
It is a very small number, but 
gets compensated by the number of gravitons in the incoming radiation field, making it feasible to observe 
quantum jumps induced in resonant mass detectors. For click detectors that monitor quantum jumps stimulated by the gravitational radiation, the signal is already at the level of $\gamma_0\Delta t \langle a^\dagger a\rangle$ as discussed. Our previous work has shown that the observable excess noise in counts is equal to $(\gamma_0\Delta t)^2 Q\langle a^\dagger a\rangle$, where $Q$ is the Mandel's $Q$ parameter~\cite{Manikandan_Wilczek_acoherence}. Thus, if the gravitational radiation field is in a quantum mechanical state for which $Q\sim O(\langle a^\dagger a\rangle)$, then the excess noise becomes observable too. We showed that a variety of plausibly relevant states, including squeezed and thermal states, have that feature.

One objective of this present work is to show that even when Mandel's $Q$ is bounded, as such is the case for sub-Poissonian states having $-1\leq Q< 0$, the quantum mechanical noise characteristics can be amplified by choosing appropriate quantum phase measurement strategies which complement click detectors.  We will show in the upcoming sections that the excess quantum noise observable in phase-quadrature measurements can be of the same order as $\gamma_0\Delta t \langle a^\dagger a\rangle$. This linear dependence on the excess noise in $\gamma_0\Delta t$ for quadrature measurements is also not surprising, as the variances of $\hat{X}$ and $\hat{P}$ have the same number (two) of creation and annihilation operators as the number operator, $a^\dagger a$. Therefore, the noise in quadrature measurements can be of the same order of magnitude as the signal in number (click) measurements.  To further substantiate this, note that for quadrature measurements, the signal is actually proportional to $\sqrt{\gamma_0\Delta t}$, the square root of the graviton-to-phonon conversion efficiency. 

More generally, we demonstrate that measurement of complementary quantum noise characteristics gives additional insight into the character of radiation fields, and can bring to light additional phenomena -- specifically, violations of the coherent state hypothesis -- that transcend classical physics. 

\section{Framework}

For later use let us briefly recall some essential results from ~\cite{Manikandan_Wilczek_acoherence}.
Given the rotating wave and single mode approximations, the problem of resonant detection of radiation fields involves the interaction Hamiltonian obeying,
\begin{equation}
    H_{I} \Delta t = \hbar\sqrt{\gamma_0\Delta t}(a^\dagger b+b^\dagger a).
\end{equation}
The mode $b$ is the detector (acoustic mode of the resonant mass detector), the mode $a$ represents the gravitational radiation field in the single mode approximation, and $\gamma_0$ is the spontaneous emission rate of the detector for the field quantum (graviton). With this model, we revisit the graviton counting statistics framework developed in Ref.~\cite{Manikandan_Wilczek_acoherence} to set the stage for the complementary phase detection strategies that follow. A generic quantum state of the field $\rho$, and vacuum of the detector $|0\rangle\langle 0|$ evolve as~\cite{Manikandan_Wilczek_acoherence},
\begin{eqnarray}
   && e^{-\frac{i}{\hbar}H \Delta t}\rho\otimes|0\rangle\langle 0|  e^{\frac{i}{\hbar}H \Delta t}\nonumber\\&=&\int d^2\alpha P(\alpha)|\alpha \cos(\sqrt{\gamma_0 \Delta t})\rangle\langle \alpha \cos(\sqrt{\gamma_0 \Delta t})|\otimes |-i\alpha \sin(\sqrt{\gamma_0\Delta t})\rangle\langle -i\alpha \sin(\sqrt{\gamma_0\Delta t})|, 
\end{eqnarray}
where we have used the Sudarshan-Glauber diagonal $P$ representation for the arbitrary state of the radiation field, $\rho$~\cite{Sudharshan,Glauber}:
\begin{equation}
    \rho = \int d^2\alpha P(\alpha)|\alpha\rangle\langle\alpha|.
\end{equation}
For click detections, we obtain the following probability, $P_{n}$ for registering $n$ clicks, given by~\cite{Manikandan_Wilczek_acoherence},
\begin{eqnarray}
   &&P_{n} =  \text{tr}_{F}\{\langle n|U_{I}(\rho\otimes |0\rangle\langle 0|) U_{I}^{\dagger}|n\rangle\}\nonumber\\&=&\frac{[\sin^2(\sqrt{\gamma_0\Delta t})]^{n}}{n!}\int d^{2}\alpha P(\alpha)|\alpha|^{2n}e^{-|\alpha|^2\sin^2(\sqrt{\gamma_0\Delta t})}.\nonumber\\
   \label{eqExact1}
\end{eqnarray}
From this, one can estimate that the average number of clicks is given by,
\begin{equation}
    \bar{n} = \sum_{n=0}^{\infty} n P_{n} =\sin^2(\sqrt{\gamma_0\Delta t}) \langle a^\dagger a\rangle \approx \gamma_0\Delta t \langle a^\dagger a\rangle ,
\end{equation}
and the variance,
\begin{equation}
    (\Delta n)^2 \approx\bar{n}+(\gamma_0\Delta t)^2 Q \langle a^\dagger a\rangle,\label{eqvar}
\end{equation}
where $Q$ is the Mandel's $Q$ parameter defined as,
\begin{eqnarray}
 Q=   \frac{\langle(\Delta \hat{N})^2\rangle -\langle \hat{N}\rangle }{\langle \hat{N}\rangle },
\end{eqnarray}
where $\hat{N} = a^{\dagger}a$. Another helpful test statistic relevant at low count rates is the ratio~\cite{Manikandan_Wilczek_acoherence}, 
\begin{equation}
    R = \frac{2P_2P_0}{P_1^2}\approx 1+Q/\langle \hat{N}\rangle .\label{ratio}
\end{equation}
For a coherent state, $Q = 0$, and it can be noted that the observable detector variance $(\Delta n)^2 = \bar{n}$ for the resonant detector. We also see that $R = 1$ for coherent states.  However, certain quantum mechanical states would show observable departures from a coherent state based on the global statistics or the ratio test.  We see that if $\gamma_0\Delta t \langle a^{\dagger}a\rangle $ is of order unity (which is satisfied by appropriate resonant bar detectors even for gravitational radiation having energy densities comparable to the LIGO-band gravitational waves~\cite{tobar_detecting_2024}), as well as if the Mandel's $Q$ parameter is of order $\langle a^{\dagger}a\rangle $, then such departures will become visible as excess quantum noise in the resonant detector~\cite{Manikandan_Wilczek_acoherence}. Our earlier work pointed out that thermal states, and highly squeezed vacuum states which are super-Poissonian, show observable deviations from coherent states based on the above acoherence criteria proposed for click detectors~\cite{Manikandan_Wilczek_acoherence}. However, the class of states that are difficult to discriminate with the above criteria are sub-Poissonian states, for which the Mandel $Q$ is bounded such that $-1\leq Q<0$. 

The smaller variance of sub-Poissonian nature for measurements in the number basis implies that appropriate quantum measurements of phase would reveal the excess noise in these quantum states in the ``phase" observable. In this present work, we explore phase-sensitive (homodyne) and phase-preserving (heterodyne) measurements to capture the excess phase noise. Through this, our goal is to propose additional tests of acoherence, that are complementary to our earlier work~\cite{Manikandan_Wilczek_acoherence}, while also probing a regime (sub-Poissonian noise) that is considered by many as the highest standard for verifying the quantum nature of radiation fields in practice.

\section{Phase sensitive quantum measurements (Homodyne detection)}

We first consider the case of phase-sensitive quantum measurements, also known as homodyne detection. For homodyne detection, the objective is to measure along one direction (given phase) in the phase-space.  This can be achieved by using a local oscillator as a phase reference to the signal at the same frequency. While an arbitrary direction can be chosen, we pick this direction as the $\hat{x}$ direction for simplicity, where the measurement operators are simply the projectors, $|x\rangle\langle x|$ on the detector, for the observable $\hat{x}=x_0 (b+b^\dagger)/\sqrt{2}:~~\hat{x}|x\rangle=x|x\rangle$. We obtain the following probabilities for observing a given quadrature $x$, given by,
\begin{eqnarray}
    P_D(x) &=& \text{Tr}_F \langle x_D| e^{-\frac{i}{\hbar}H \Delta t}\rho\otimes|0\rangle\langle 0|  e^{\frac{i}{\hbar}H \Delta t}|x_D\rangle \nonumber\\
    &=&\frac{1}{\sqrt{\pi}x_0}\int d^2\alpha P(\alpha) e^{-\frac{[x-\sqrt{2}x_0 \text{Im}(\alpha)\sin(\sqrt{\gamma_0\Delta t})]^2}{x_0^2}},
\end{eqnarray}
where $x_0 \propto\sqrt{\hbar}$ is the zero-point length of the detector mode. Here, we have also made use of the relation, $\text{Re}[-i\alpha\sin(\sqrt{\gamma_0\Delta t})]=\text{Im}(\alpha)\sin(\sqrt{\gamma_0\Delta t}).$ We find that,
\begin{equation}
    \langle \hat{x}\rangle = \sqrt{2}x_0\sin(\sqrt{\gamma_0\Delta t}) \int d^2\alpha  P(\alpha)\text{Im}(\alpha) = \sqrt{2}x_0\sin(\sqrt{\gamma_0\Delta t})\langle \text{Im}(\alpha)\rangle.
\end{equation}
Similarly, we have,
\begin{equation}
    \langle \hat{x}^2\rangle = \frac{1}{2}x_0^2 +2x_0^2\sin^2(\sqrt{\gamma_0\Delta t}) \int d^2\alpha  P(\alpha)\text{Im}(\alpha)^2 =  \frac{1}{2}x_0^2 +2x_0^2\sin^2(\sqrt{\gamma_0\Delta t})  \langle\text{Im}(\alpha)^2\rangle.
\end{equation}
We obtain that the variance is given by,
\begin{eqnarray}
    \langle (\Delta \hat{x})^2\rangle &=&\langle \hat{x}^2\rangle-\langle \hat{x}\rangle^2 = \frac{1}{2}x_0^2 +2x_0^2\sin^2(\sqrt{\gamma_0\Delta t})  [\langle\text{Im}(\alpha)^2\rangle-\langle \text{Im}(\alpha)\rangle^2]\nonumber\\
    &=&x_0^2\bigg[\frac{1}{2}+\sin^2(\sqrt{\gamma_0\Delta t})\bigg(\langle (\Delta \hat{P})^2\rangle -\frac{1}{2}\bigg)\bigg].
\end{eqnarray}
We have made use of the result, 
\begin{equation}
   2 [\langle\text{Im}(\alpha)^2\rangle-\langle \text{Im}(\alpha)\rangle^2] = \langle (\Delta \hat{P})^2\rangle -\frac{1}{2},
\end{equation}
where $\hat{P}=(a-a^\dagger)/(i\sqrt{2}),~\hat{X}=(a+a^\dagger)/\sqrt{2} $ are the quadratures of the field represented in dimensionless units (see Appendix.~\ref{appA}). We can now estimate the observable quantum noise for different quantum states of the radiation field. First we estimate the noise observable for a coherent state $\rho = |\alpha\rangle\langle\alpha|$ for which $\langle (\Delta \hat{P})^2\rangle =1/2.$ We find that,
\begin{equation}
     \langle (\Delta \hat{x})^2\rangle=\frac{x_0^2}{2}.
\end{equation}
For a thermal state of the gravitational radiation with average thermal quanta $n_{th}$, the $P$ function, and the momentum variance are given by,
\begin{equation}
    P_{th}(\alpha)=\frac{1}{\pi n_{th}}e^{-|\alpha|^2/n_{th}},~~\langle (\Delta \hat{P})^2\rangle=\frac{2n_{th}+1}{2}.
\end{equation}
We obtain,
    \begin{eqnarray}
    \langle (\Delta \hat{x})^2\rangle &=& \frac{1}{2}x_0^2 +2x_0^2\sin^2(\sqrt{\gamma_0\Delta t})  [\langle\text{Im}(\alpha)^2\rangle-\langle \text{Im}(\alpha)\rangle^2]\nonumber\\
    &=& \frac{1}{2}x_0^2 +2x_0^2\sin^2(\sqrt{\gamma_0\Delta t})\frac{n_{th}}{2} = x_0^2 \bigg[n_{th}\sin^2(\sqrt{\gamma_0\Delta t})+\frac{1}{2}\bigg].
\end{eqnarray}
If the number of thermal gravitons are of the same order that the resonant detector would register a click: $\langle a^\dagger a\rangle \sin^2(\sqrt{\gamma_0\Delta t})\approx \gamma_0\Delta t \langle a^\dagger a\rangle \sim O(1)$, we see that the quantum noise of gravitons in the thermal state also show a similar order of magnitude difference in the excess noise in homodyne measurements. For Weber bars used as click detectors for gravitons as proposed in Ref.~\cite{tobar_detecting_2024}, stimulated absorption probability $\gamma_0\langle a^\dagger a\rangle\Delta t$ of order unity can be achieved for gravitational radiation with energy densities comparable to the LIGO band. Hence it appears that such tests are feasible within reasonable time windows ($\Delta t\sim 1$ms). While click detection strategy was proposed only recently, homodyne detection has the added advantage that it is similar in spirit to how Weber's classical detection strategy would have worked.

We can also generalize our results for a generic Gaussian state (a displaced, squeezed, thermal state. See Appendix.~\ref{appA}),
\begin{equation}
    \langle (\Delta \hat{x})^2\rangle =x_0^2 \bigg[\bigg\{\frac{2n_{th}+1}{2}[\cosh(2r)+\sinh(2r)\cos(\phi)]-\frac{1}{2}\bigg\}\sin^2(\sqrt{\gamma_0\Delta t})+\frac{1}{2}\bigg].
\end{equation}
Consider gravitational radiation in a squeezed vacuum state: $n_{th} = 0$ and $\phi = 0$. In this case, we get,
\begin{equation}
    \langle (\Delta \hat{x})^2\rangle =x_0^2 \bigg[\bigg\{\frac{1}{2}\exp(2r)-\frac{1}{2}\bigg\}\sin^2(\sqrt{\gamma_0\Delta t})+\frac{1}{2}\bigg]\rightarrow  \frac{x_0^2}{2}\bigg[\cos^2(\sqrt{\gamma_0\Delta t})+\exp(2r)\sin^2(\sqrt{\gamma_0\Delta t})\bigg],
\end{equation}
which again suggests that excess quantum noise along the other quadrature is observable, since $\exp(2r)\sin^2(\sqrt{\gamma_0\Delta t})\sim \exp(2r)\gamma_0\Delta t\sim O(1)$ for large squeezing such that $\langle a^\dagger a\rangle  = \sinh(r)^2\sim \exp(2r)\sim 10^{36}$, which corresponds to number of gravitons with energy densities comparable to LIGO band gravitational waves.

In the special case $\phi=\pi$ one obtains, 
\begin{equation}
    \langle (\Delta \hat{x})^2\rangle =x_0^2 \bigg[\bigg\{\frac{1}{2}\exp(-2r)-\frac{1}{2}\bigg\}\sin^2(\sqrt{\gamma_0\Delta t})+\frac{1}{2}\bigg],
\end{equation}
Then in the limit $r\rightarrow\infty$ we find
\begin{equation}
    \langle (\Delta \hat{x})^2\rangle =x_0^2 \bigg[\bigg\{\frac{1}{2}\exp(-2r)-\frac{1}{2}\bigg\}\sin^2(\sqrt{\gamma_0\Delta t})+\frac{1}{2}\bigg]\rightarrow  \frac{x_0^2}{2}\bigg[1-\sin^2(\sqrt{\gamma_0\Delta t})\bigg],
\end{equation}
Here
noise below vacuum noise is difficult to measure, since $\gamma_0\Delta t$ alone is very small. This difficulty was also pointed out recently in~\cite{carney_comments_2024,CarneyPRD}.

It is interesting to see what results we find for a Fock (number) state that is maximally sub-Poissonian. For a number state $|n\rangle$ we know that, 
\begin{eqnarray}
    \langle (\Delta \hat{P})^2 
\rangle= \langle (\Delta \hat{X})^2 \rangle=\frac{2n+1}{2}. 
\end{eqnarray}
Hence, we obtain that,
\begin{eqnarray}
    \langle (\Delta \hat{x})^2\rangle &=&x_0^2\bigg[\frac{1}{2}+\sin^2(\sqrt{\gamma_0\Delta t})n\bigg].
\end{eqnarray}
If the number of gravitons making up the gravitational radiation in this context has energy densities comparable to that of LIGO band gravitational waves ($n \sim 10^{36}$), we can have $n \sin^2(\sqrt{\gamma_0\Delta t})\approx n\gamma_0\Delta t$ of order unity for typical Weber bar detectors. This suggests that for our phase-sensitive quantum measurement strategy, the deviation in the excess noise produced when compared to a coherent state of the radiation field can be of order unity. 
This should be compared to statistical noise in direct click detection, which does not discriminate between Fock and coherent states ~\cite{Manikandan_Wilczek_acoherence}. 

While our tests are based on the excess quantum noise produced (i.e., in the second moment), it can be noted that the average signal also exhibit complementary features depending on the quantum state of the radiation field, and the detection strategy used~\cite{loughlin_wave-particle_2025}.  

\section{Heterodyne detection\label{AppB}}

Heterodyne detection corresponds to projecting the field on to the coherent state basis, described by measurement operators,
\begin{equation}
  K_{\beta} =  \frac{1}{\sqrt{\pi}}|\beta\rangle\langle \beta|.
\end{equation}
The factor of $\pi$ comes from the overcompleteness of coherent states. For the interaction Hamiltonian we have considered, the probability of measuring a given coherent state $|\beta\rangle$ is given by,
\begin{eqnarray}
    P_D(\beta) &=& \frac{1}{\pi}\text{Tr}_F \langle \beta_D| e^{-\frac{i}{\hbar}H \Delta t}\rho\otimes|0\rangle\langle 0|  e^{\frac{i}{\hbar}H \Delta t}|\beta_D\rangle \nonumber\\
    &=&\frac{1}{\pi}\int d^2\alpha P(\alpha) e^{-|\beta +i\alpha \sin(\sqrt{\gamma_0\Delta t})|^2}.\label{pdbeta}
\end{eqnarray}

\subsection{Statistics of quadratures}
From $P_D(\beta)$ given in Eq.~\eqref{pdbeta}, we can estimate the moments of measured quadrature outcomes,
\begin{eqnarray}
    \langle \text{Re}(\beta)\rangle  &=& \int d^2\beta \text{Re}(\beta)\frac{1}{\pi}\int d^2\alpha P(\alpha) e^{-|\beta +i\alpha \sin(\sqrt{\gamma_0\Delta t})|^2}\nonumber\\
    &=&\sin(\sqrt{\gamma_0\Delta t})\int d^2\alpha P(\alpha)\text{Im}(\alpha) = \sin(\sqrt{\gamma_0\Delta t})\langle \text{Im}(\alpha)\rangle.\nonumber\\
    \langle \text{Re}(\beta)^2\rangle&=&\frac{1}{2}+\sin^2(\sqrt{\gamma_0\Delta t})\int d^2\alpha P(\alpha)\text{Im}(\alpha)^2=\frac{1}{2}+\sin^2(\sqrt{\gamma_0\Delta t})\langle \text{Im}(\alpha)^2\rangle.\nonumber\\
        \langle \text{Im}(\beta)\rangle  &=& \int d^2\beta \text{Re}(\beta)\frac{1}{\pi}\int d^2\alpha P(\alpha) e^{-|\beta +i\alpha \sin(\sqrt{\gamma_0\Delta t})|^2}\nonumber\\
    &=&-\sin(\sqrt{\gamma_0\Delta t})\int d^2\alpha P(\alpha)\text{Re}(\alpha) = -\sin(\sqrt{\gamma_0\Delta t})\langle \text{Re}(\alpha)\rangle.\nonumber\\
    \langle \text{Im}(\beta)^2\rangle&=&\frac{1}{2}+\sin^2(\sqrt{\gamma_0\Delta t})\int d^2\alpha P(\alpha)\text{Re}(\alpha)^2=\frac{1}{2}+\sin^2(\sqrt{\gamma_0\Delta t})\langle \text{Re}(\alpha)^2\rangle.
\end{eqnarray}
These results yield the following estimates for the variances,
\begin{eqnarray}
    \langle(\Delta \text{Re}(\beta))^2\rangle &=& \frac{1}{2}+\sin^2(\sqrt{\gamma_0\Delta t})[\langle \text{Im}(\alpha)^2\rangle-\langle \text{Im}(\alpha)\rangle^2]\nonumber\\
    &=&\frac{1}{2}\bigg\{1+\sin^2(\sqrt{\gamma_0\Delta t})\bigg[\langle (\Delta \hat{P})^2\rangle-\frac{1}{2}\bigg]\bigg\},
\end{eqnarray}
and
\begin{eqnarray}
    \langle(\Delta \text{Im}(\beta))^2\rangle &=& \frac{1}{2}+\sin^2(\sqrt{\gamma_0\Delta t})[\langle \text{Re}(\alpha)^2\rangle-\langle \text{Re}(\alpha)\rangle^2]\nonumber\\
    &=&\frac{1}{2}\bigg\{1+\sin^2(\sqrt{\gamma_0\Delta t})\bigg[\langle (\Delta \hat{X})^2\rangle-\frac{1}{2}\bigg]\bigg\}.
\end{eqnarray}
As an example, we consider a squeezed state with $\phi = 0$ and $n_{th} = 0$. In this case, $\langle (\Delta \hat{P})^2\rangle\rightarrow e^{2r}/2$, and $\langle (\Delta \hat{X})^2\rangle\rightarrow e^{-2r}/2$. In the limit of large squeezing ($r\rightarrow\infty$) we find that,
\begin{eqnarray}
    \langle(\Delta \text{Im}(\beta))^2\rangle &=&\frac{1}{2}\bigg\{1+\sin^2(\sqrt{\gamma_0\Delta t})\bigg[\langle (\Delta \hat{X})^2\rangle-\frac{1}{2}\bigg]\bigg\}\nonumber\\
    &=&\frac{1}{2}\bigg\{1+\sin^2(\sqrt{\gamma_0\Delta t})\bigg[\frac{e^{-2r}}{2}-\frac{1}{2}\bigg]\bigg\}\rightarrow \frac{1}{2}\cos^2(\sqrt{\gamma_0\Delta t})\approx \frac{1}{2},
\end{eqnarray}
and the last equality holds since $\gamma_0\Delta t\ll 1$. However we can consider the noise observable in the conjugate signal. For $r\gg 1$, we find,
\begin{eqnarray}
    \langle(\Delta \text{Re}(\beta))^2\rangle &=&\frac{1}{2}\bigg\{1+\sin^2(\sqrt{\gamma_0\Delta t})\bigg[\langle (\Delta \hat{P})^2\rangle-\frac{1}{2}\bigg]\bigg\}\nonumber\\
    &=&\frac{1}{2}\bigg\{1+\sin^2(\sqrt{\gamma_0\Delta t})\bigg[\frac{e^{2r}}{2}-\frac{1}{2}\bigg]\bigg\}\rightarrow \bigg[\frac{1}{2}\cos^2(\sqrt{\gamma_0\Delta t})+\frac{e^{2r}}{4}\sin^2(\sqrt{\gamma_0\Delta t})\bigg].
\end{eqnarray}
This again suggests that a large squeezing can make the excess noise of gravitons visible as noise in one of the quadratures. However, note that the excess noise observable in each quadrature of heterodyne is a factor of two smaller than that observable in the homodyne signal (in units where we set $x_0=1$).

Next we also consider the detector response to heterodyne quadrature measurements for a maximally sub-Poissonian number state of the radiation field. We obtain,
\begin{eqnarray}
    \langle(\Delta \text{Re}(\beta))^2\rangle &=& \frac{1}{2}+\sin^2(\sqrt{\gamma_0\Delta t})[\langle \text{Im}(\alpha)^2\rangle-\langle \text{Im}(\alpha)\rangle^2]\nonumber\\
    &=&\frac{1}{2}\bigg\{1+\sin^2(\sqrt{\gamma_0\Delta t})\bigg[\langle (\Delta \hat{P})^2\rangle-\frac{1}{2}\bigg]\bigg\}\nonumber\\
    &=&\frac{1}{2}\bigg\{1+\sin^2(\sqrt{\gamma_0\Delta t})n\bigg\}
\end{eqnarray}
and
\begin{eqnarray}
    \langle(\Delta \text{Im}(\beta))^2\rangle &=& \frac{1}{2}+\sin^2(\sqrt{\gamma_0\Delta t})[\langle \text{Re}(\alpha)^2\rangle-\langle \text{Re}(\alpha)\rangle^2]\nonumber\\
    &=&\frac{1}{2}\bigg\{1+\sin^2(\sqrt{\gamma_0\Delta t})\bigg[\langle (\Delta \hat{X})^2\rangle-\frac{1}{2}\bigg]\bigg\}\nonumber\\
    &=&\frac{1}{2}\bigg\{1+\sin^2(\sqrt{\gamma_0\Delta t})n\bigg\}
\end{eqnarray}
We arrive at the same conclusion as before. The noise characteristics in the heterodyne signal for a Fock state are similar to those of a thermal state, a factor of two smaller when compared to homodyne. Nevertheless, as is evident, both homodyne and heterodyne detection strategies can discriminate a Fock state from a coherent state, while click detectors cannot. Given that click detectors can discriminate a thermal state from a coherent state, these different (phase and number) quantum measurement strategies can be combined to definitively test the acoherence of gravitational radiaiton through the methods we have discussed here.

Finally, we note that the measurements of power fluctuations in the heterodyne signal tells a different story, that is comparable to click detectors. In Sec.~\ref{power}, we show that measuring heterodyne power fluctuations does not allow to discriminate the number state from a coherent state, while thermal states are discriminable. This also suggests that combining different measurement strategies, it is possible to map out the quantum states of the radiation field, including that for gravitational radiation using bar detectors.

\subsection{Statistics of power in the heterodyne signal\label{power}}
We first estimate the moments of the heterodyne power, which we define as, $\hat{J}=b^\dagger b$. For the average, we obtain,
\begin{eqnarray}
    \langle \hat{J}\rangle &=& \langle b^\dagger b\rangle_D =\int d^2\beta P_D(\beta)|\beta|^2 
    =\frac{1}{\pi}\int d^2\alpha P(\alpha)\int d^2\beta |\beta|^2  e^{-|\beta +i\alpha \sin(\sqrt{\gamma_0\Delta t})|^2}\nonumber\\
    &=&\int d^2\alpha P(\alpha)[1+\sin^2(\sqrt{\gamma_0\Delta t})|\alpha|^2]=1+\sin^2(\sqrt{\gamma_0\Delta t})\langle a^\dagger a\rangle .
\end{eqnarray}

Similarly, we can write the variance as,
\begin{eqnarray}
    \langle (\Delta \hat{J})^2\rangle  &=&\langle (b^\dagger b)^2\rangle -\langle b^\dagger b\rangle ^2 =\langle b^\dagger b\rangle+\langle (b^\dagger)^2 b^2\rangle-\langle b^\dagger b\rangle^2\nonumber\\&=&\langle |\beta|^2\rangle_D+\langle |\beta|^4\rangle_D-\langle |\beta|^2\rangle_D^2\nonumber\\
    &=&\int d^2\beta P_D(\beta)[|\beta|^2+|\beta|^4-\langle |\beta|^2\rangle^2 ]\nonumber\\
    &=&\frac{1}{\pi}\int d^2\alpha P(\alpha)\int d^2\beta  e^{-|\beta +i\alpha \sin(\sqrt{\gamma_0\Delta t})|^2}[|\beta|^2+|\beta|^4-\langle |\beta|^2\rangle^2 ].
\end{eqnarray}
We evaluated the first term above already, and the third term is simply the square of the first integral since it is integrating a constant (squared average). The second term gives,
\begin{eqnarray}
    \frac{1}{\pi}\int d^2\alpha P(\alpha)\int d^2\beta  e^{-|\beta +i\alpha \sin(\sqrt{\gamma_0\Delta t})|^2}|\beta|^4 &=& \int d^2\alpha P(\alpha) [2+4|\alpha|^2\sin^2(\sqrt{\gamma_0\Delta t})+|\alpha|^4\sin^4(\sqrt{\gamma_0\Delta t})] \nonumber\\
    &=& 2+ 4 \sin^2(\sqrt{\gamma_0\Delta t})\langle a^\dagger a\rangle +\sin^4(\sqrt{\gamma_0\Delta t})\langle (a^\dagger)^2 a^2\rangle .
\end{eqnarray}
We can substitute this back, and find,
\begin{eqnarray}
     \langle (\Delta \hat{J})^2\rangle  &=& \frac{1}{\pi}\int d^2\alpha P(\alpha)\int d^2\beta  e^{-|\beta +i\alpha \sin(\sqrt{\gamma_0\Delta t})|^2}[|\beta|^2+|\beta|^4-\langle |\beta|^2\rangle^2 ]\nonumber\\
     &=& 2+ 3 \sin^2(\sqrt{\gamma_0\Delta t})\langle a^\dagger a\rangle +\sin^4(\sqrt{\gamma_0\Delta t})[\langle (a^\dagger)^2 a^2\rangle -\langle a^\dagger a\rangle ^2]\nonumber\\
     &=&2+ 3 \sin^2(\sqrt{\gamma_0\Delta t})\langle a^\dagger a\rangle +\sin^4(\sqrt{\gamma_0\Delta t})[\langle (a^\dagger a)^2-a^\dagger a\rangle -\langle a^\dagger a\rangle ^2]\nonumber\\
     &=&2+ 3 \sin^2(\sqrt{\gamma_0\Delta t})\langle \hat{N}\rangle +\sin^4(\sqrt{\gamma_0\Delta t})[\langle (\Delta \hat{N})^2\rangle -\langle \hat{N}\rangle ]\nonumber\\
     &=&2+ 3 \sin^2(\sqrt{\gamma_0\Delta t})\langle \hat{N}\rangle +\sin^4(\sqrt{\gamma_0\Delta t})Q\langle \hat{N}\rangle,
\end{eqnarray}
where $\hat{N}=a^\dagger a$. For $\gamma_0\Delta t\ll 1$, we can write $\sin^2(\sqrt{\gamma_0\Delta t})\approx \gamma_0\Delta t$. In this case, we have,
\begin{equation}
    \langle \hat{J}\rangle\approx 1+\gamma_0\Delta t\langle \hat{N}\rangle ,~~\text{and},~~\langle (\Delta \hat{J})^2\rangle \approx 2+ 3 \gamma_0\Delta t\langle \hat{N}\rangle +(\gamma_0\Delta t)^2Q\langle \hat{N}\rangle  = (3\langle \hat{J}\rangle -1)+(\gamma_0\Delta t)^2Q\langle \hat{N}\rangle .
\end{equation}
We see that super-Poissonian states which have a large $Q$ can make the noise of the radiation field measurable in heterodyne power fluctuations too. In the gravitational context, we have $\langle \hat{N}\rangle \sin^2(\sqrt{\gamma_0\Delta t})\approx \gamma_0\Delta t \langle \hat{N}\rangle \sim O(1)$ for Weber bars ($\gamma_0\sim 10^{-33}$Hz), and for intense ($\langle \hat{N}\rangle \sim 10^{36}$) gravitational radiation with energy densities comparable to LIGO band graviational waves, within reasonable time windows ($\Delta t\sim 1$ms). For $Q\sim O(\langle \hat{N}\rangle )$, we will also be operating in a regime where $(\gamma_0\Delta t)^2Q\langle \hat{N}\rangle $ is substantial.  Hence the excess quantum noise of gravitons in these states will manifest as excess noise in the power of the heterodyne signal. However, as shown above, the noise characteristics, although differ in form from what a number detector could measure as described in Ref.~\cite{Manikandan_Wilczek_acoherence}, have a similar nature.

Finally, note that above we defined the heterodyne power fluctuations based on the operator $\hat{J}=b^\dagger b$. Alternatively one could think of a classical signal $j=|\beta|^2$ and estimate its fluctuations, $(\Delta j)^2=\langle j^2\rangle - \langle j\rangle^2$. Note the measured fluctuations will therefore be different. We obtain that mean is same as before, $\langle j\rangle  \approx 1+\gamma_0\Delta t\langle a^\dagger a\rangle $, and that the variance  $\langle (\Delta j)^2\rangle = (2\langle j\rangle -1)+(\gamma_0\Delta t)^2 Q\langle a^\dagger a\rangle.$

\section{Conclusion}

We have compared different quantum measurement strategies: the measurements of quantum jumps (clicks), and diffusive quantum phase measurements, and discussed how they can be complementary for probing the quantum mechanical character of gravitational radiation using resonant mass detectors. The feasibility of such tests is justified by recent observations that resonant mass detectors allow us to have stimulated absorption probabilities that are of order unity for gravitational radiation with chirp frequencies and energy densities comparable to the LIGO band~\cite{tobar_detecting_2024}.   Our analysis suggests that the excess quantum mechanical noise in different quantum mechanical states of the radiation field must manifest in a complementary manner when subjected to different (number or phase) quantum measurement detection strategies. In particular, the sub-Poissonian quantum character in gravitational radiation, if it exists, may also be observable as excess noise in a phase-sensitive measurement, while it may remain elusive to click detectors. Our proposed tests can therefore be used to definitively address the question of whether gravitational radiation reveals its quantum character when measured. Transduction of the signal to smaller masses may also be possible~\cite{tobar_detecting_2025}, which may further enhance the feasibility of the tests proposed here.

Sources for gravitational radiation with quantum mechanical characteristics are also likely to exist in nature. 
Indeed, quantum effects such as squeezing require nonlinearities at the source, and general relativity is an inherently nonlinear theory. Single- and multi-mode squeezing of gravitational radiation has been proposed in some contexts.  For instance, cosmological expansion is one paradigm that is expected to squeeze primordial gravitational radiation~\cite{grishchuk_quantum_1989,GrishchukPRD}. The simplest of such multi-mode scenarios we could consider is pair-creation in the gravitational context where gravitons are produced in pairs via interactions of the type,
\begin{equation}
    V_{I} \sim -ig(a_{k}a_{-k}-a_{k}^\dagger a_{-k}^\dagger).
\end{equation}
 Above $g$ can be thought of as the pump or source mode in the semiclassical approximation, analogous to the pump mode in parametric down conversion. The resultant quantum state of gravitons will be the two-mode squeezed vacuum state,
 \begin{equation}
     |\psi\rangle = \exp{(-iV_I t)}|0\rangle_k|0\rangle_{-k}=\sum_{k}\frac{[\tanh(gt)]^n}{\cosh(gt)}|n\rangle_k |n\rangle_{-k}.
 \end{equation}
We can use this quantum state to predict the results for a ratio test that is possible with a click detector. Assume that our detector only has access to one of the modes from the pair, say, with wave vector $k$. To the leading order, we find the following click probabilities, $P_0\approx 1$, $P_1 = \gamma_0\Delta t \langle a_{k}^\dagger a_k\rangle$, and $P_2\approx (\gamma_0\Delta t)^2 \langle (a_{k}^\dagger)^2 a^2_k\rangle/2$. Since $\langle a_{k}^\dagger a_k\rangle=\sinh^2(gt)$ and $\langle (a_{k}^\dagger)^2 a^2_k\rangle = \sinh^4(gt)$ for the quantum state $|\psi\rangle$, we obtain the ratio $R=2P_0P_2/P_1^2 = 2$. The result is different from the coherent state result ($R=1$), and is equivalent to the result expected for a thermal state with average quanta $n_{th} = \sinh^2(gt)$. It is not surprising, as tracing out one of the modes from a two-mode squeezed vacuum state results in a thermal state for the other mode.

While a systematic analysis of nonlinearities involved can be substantially more challenging close to the merger, it has also been suggested that head-on collisions of black holes may also generate acoherent quantum mechanical states of this nature~\cite{lovas_quantization_2001}. Hence, it is highly suggestive that the non-classical features of the gravitational radiation can be probed in near-future resonant bar detector-based measurements using the methods we have proposed here. The quantum noise characteristics observable in ground-based resonant mass detectors could reveal important information about various quantum mechanical effects that are relevant at the sources of gravitational radiation and in propagation, thereby offering a new quantum mechanical window into the observable universe.

\section{Acknowledgments}
FW is supported by the U.S. Department of Energy under grant Contract Number DE-SC0012567 and by the Swedish Research Council under Contract No. 335-2014-7424. SKM is supported in part by the Swedish Research Council under Contract No. 335-2014-7424 and in part by the Wallenberg Initiative on Networks and Quantum Information (WINQ).  We thank Maulik Parikh and Igor Pikovski for stimulating conversations around these subjects.  

\bigskip
\noindent\textit{Note} An essay summarizing the results presented here was submitted to the Gravity Research Foundation (GRF) essay competition on 31st March, 2025~\cite{Essay}.

\bigskip
\appendix
\begin{widetext}
\section{Summary of useful results\label{appA}}
Using dimensionless representation for $\hat{X}$ and $\hat{P}$:
\begin{equation}
\hat{X}= \frac{1}{\sqrt{2}}(a+a^\dagger),~~\hat{P} =\frac{1}{\sqrt{2}i}(a-a^\dagger), 
\end{equation} we obtain, $\langle\alpha| \hat{X}|\alpha\rangle =\sqrt{2}\text{Re}(\alpha)$ and $\langle\alpha| \hat{X}^2|\alpha\rangle =\text{Re}(\alpha^2)+|\alpha|^2+1/2=2 \text{Re}(\alpha)^2+1/2$. Similarly, $\langle\alpha| \hat{P}|\alpha\rangle =\sqrt{2}\text{Im}(\alpha)$ and $\langle\alpha| \hat{P}^2|\alpha\rangle =-\text{Re}(\alpha^2)+|\alpha|^2+1/2=2 \text{Im}(\alpha)^2+1/2$. We can use these results to estimate the position and momentum variances in the P representation for an arbitrary density matrix:
\begin{equation}
    \rho = \int d^2\alpha P(\alpha)|\alpha\rangle\langle\alpha|.
\end{equation}
From this we find,
\begin{eqnarray}
    \langle(\Delta \hat{X})^2\rangle  &=& \frac{1}{2}+2\int d^2\alpha P(\alpha) \text{Re}(\alpha)^2 -2\bigg[\int d^2\alpha P(\alpha) \text{Re}(\alpha)\bigg]^2=\frac{1}{2}+2\langle(\Delta\text{Re}(\alpha))^2\rangle,\nonumber\\
    \langle(\Delta \hat{P})^2\rangle  &=& \frac{1}{2}+2\int d^2\alpha P(\alpha) \text{Im}(\alpha)^2 -2\bigg[\int d^2\alpha P(\alpha) \text{Im}(\alpha)\bigg]^2=\frac{1}{2}+2\langle(\Delta\text{Im}(\alpha))^2\rangle.
\end{eqnarray}
This yields,
\begin{eqnarray}
    \langle(\Delta\text{Im}(\alpha))^2\rangle&=&\frac{1}{2}\bigg[\langle(\Delta \hat{P})^2\rangle -\frac{1}{2}\bigg],~~\text{and}\nonumber\\
    \langle(\Delta\text{Re}(\alpha))^2\rangle&=&\frac{1}{2}\bigg[\langle(\Delta \hat{X})^2\rangle -\frac{1}{2}\bigg].
\end{eqnarray}
For a generic Gaussian state, we can use,
\begin{eqnarray}
    \langle (\Delta \hat{P})^2\rangle &=& \frac{2n_{th}+1}{2}[\cosh(2r)+\sinh(2r)\cos(\phi)],
\end{eqnarray}
and 
\begin{eqnarray}
    \langle (\Delta \hat{X})^2\rangle &=& \frac{2n_{th}+1}{2}[\cosh(2r)-\sinh(2r)\cos(\phi)].
\end{eqnarray}It is easy to see that the above results are true for a coherent state and a thermal state. As for a non-trivial check, we may use the following $P$ function used for a squeezed vacuum state~\cite{schleich_quantum_2011,leuchs_intensityintensity_2015},
\begin{eqnarray}
    P(\alpha) = [e^{O_R} \delta (\alpha_R)][e^{-O_I} \delta (\alpha_I)],
\end{eqnarray}
where,
\begin{eqnarray}
    O_r = \frac{1-e^{2r}}{8 e^{2r}}\frac{\partial^2}{\partial\alpha_R^2},
\end{eqnarray}
and
\begin{eqnarray}
    O_I = \frac{1-e^{2r}}{8}\frac{\partial^2}{\partial\alpha_I^2}.
\end{eqnarray}
We have also denoted Re/Im($\alpha$)$=\alpha_{R/I}$. From this, we can estimate the mean, and the variances. Note that, for the above $P$ function, we can easily verify that the mean is equal to zero. For the variance,
\begin{equation}
  \int  d^2\alpha \alpha_R^2P(\alpha) = \int d\alpha_R (\alpha_R)^2 [e^{O_R} \delta (\alpha_R)] = \int d\alpha_R \delta (\alpha_R) [e^{O_R} (\alpha_R)^2]=\frac{1-e^{2r}}{4 e^{2r}}.
\end{equation}
similarly,
\begin{equation}
  \int  d^2\alpha \alpha_I^2P(\alpha) = \int d\alpha_I (\alpha_I)^2 [e^{-O_I} \delta (\alpha_I)] = \int d\alpha_I \delta (\alpha_I) [e^{-O_I} (\alpha_I)^2]=\frac{e^{2r}-1}{4}.
\end{equation}
From these we find,
\begin{eqnarray}
   \langle(\Delta \hat{X})^2\rangle = \frac{1}{2}+2\langle(\Delta\text{Re}(\alpha))^2\rangle&=&e^{-2r}/2,
\end{eqnarray}
and
\begin{eqnarray}
  \langle(\Delta \hat{P})^2\rangle =  \frac{1}{2}+2\langle(\Delta\text{Im}(\alpha))^2\rangle&=&e^{2r}/2.
\end{eqnarray}
The result matches explicitly.

\end{widetext}
\bibliography{ref}
\end{document}